\documentclass[conference]{IEEEtran}
\IEEEoverridecommandlockouts

\def\BibTeX{{\rm B\kern-.05em{\sc i\kern-.025em b}\kern-.08em
    T\kern-.1667em\lower.7ex\hbox{E}\kern-.125emX}}

\usepackage{url}
\usepackage{ifthen}
\usepackage{cite}
\usepackage[cmex10]{amsmath} 

\usepackage{graphicx}
\usepackage[caption=false, font=footnotesize]{subfig}
\usepackage{comment}

\usepackage{algorithmicx}        
\usepackage{algorithm}          
\usepackage{algpseudocode}
\usepackage{bbm}

\usepackage{tabularx, booktabs}

\algdef{SE}[DOWHILE]{Do}{doWhile}{\algorithmicdo}[1]{\algorithmicwhile\ #1}%

\usepackage{amsmath}
\usepackage{amssymb}
\usepackage{xcolor}

\newtheorem{theorem}{\bfseries Theorem}
\newtheorem{lemma}{\bfseries Lemma}
\newtheorem{definition}{\bfseries Definition}

\newtheorem{remark}{\bfseries Remark}

\newtheorem{example}{\bfseries Example}
\newtheorem{assumption}{\bfseries Assumption}

\DeclareMathOperator*{\argmin}{arg\,min}
\DeclareMathOperator{\Imax}{\bar{I}}

\begin{document}

\title{On Distributionally Robust Lossy Source Coding}

\author{
    \IEEEauthorblockN{ Giuseppe Serra$^{\dagger}$, Photios A. Stavrou$^{\dagger}$, Marios Kountouris$^{\dagger}$$^{\ddagger}$} \IEEEauthorblockA{
    $^{\dagger}$Communication Systems Department, EURECOM, Sophia-Antipolis, France\\
    $^{\ddagger}$DaSCI, Department of Computer Science and AI, University of Granada, Spain\\
    Email: \{giuseppe.serra, fotios.stavrou\}@eurecom.fr, mariosk@ugr.es}
    \thanks{ The work of G. Serra and M. Kountouris is supported by the European Research Council (ERC) under the European Union’s Horizon 2020 Research and Innovation Programme (Grant agreement No. 101003431). The work of P. A. Stavrou is supported in part by the SNS JU project 6G-GOALS under the EU’s Horizon programme (Grant Agreement No. 101139232) and by the Huawei France-EURECOM Chair on Future Wireless Networks.}
  }

\maketitle

\begin{abstract} 
In this paper, we investigate the problem of distributionally robust source coding, i.e., source coding under uncertainty in the source distribution, discussing both the coding and computational aspects of the problem. We propose two extensions of the so-called \textit{Strong Functional Representation Lemma} (SFRL), considering the cases where, for a fixed conditional distribution, the marginal inducing the joint coupling belongs to either a finite set of distributions or a Kullback–Leibler divergence sphere (KL-Sphere) centered at a fixed nominal distribution. Using these extensions, we derive distributionally robust coding schemes for both the one-shot and asymptotic regimes, generalizing previous results in the literature. Focusing on the case where the source distribution belongs to a given KL-Sphere, we derive an implicit characterization of the points attaining the robust rate-distortion function (R-RDF), which we later exploit to implement a novel algorithm for computing the R-RDF. Finally, we characterize the analytical expression of the R-RDF for Bernoulli sources, providing a theoretical benchmark to evaluate the estimation performance of the proposed algorithm. 
\end{abstract}

\IEEEpeerreviewmaketitle

\section{Introduction}

Distributionally robust source coding considers the fundamental problem of source coding under uncertainty on the source statistics, i.e., the real source statistic is \textit{unknown} but assumed to belong to a set of possible candidates. The relevancy of this scenario is shown in many real-world applications, e.g., compression of IoT sensor data in variable conditions \cite{IoTShift}, distributional resilient deep neural network codecs \cite{shift:chen:2023toward}, where the ambiguity of the information source statistic may heavily impact the coding process. Universal variable-length codes that adjust their encoding rate based on the observed source statistic have been a long-standing focus of research in information theory. In the context of lossless compression, Lempel-Ziv codes \cite{Ziv:1977,ziv:1978} asymptotically achieve the entropy rate by implementing a mechanism that estimates the source statistic from the observed data. Similar ideas have also been employed to some degree in the context of lossy compression, see e.g. \cite{yang:1996, Yang:1997}. However, in many applications, the worst-case performance over a class of possible sources is a more representative description of the operative conditions. In this setting, the fundamental compression limit is characterized as a zero-sum game \cite{washburn:2014two} between the designer choosing a coding strategy and nature selecting the worst possible source distribution given the chosen code. This idea is expressed as a min-max optimization problem, known as the compound or robust rate-distortion function (R-RDF) \cite{SAKRISON:1969165}. Depending on the considered class of sources, various characterizations of the R-RDF have been studied in the literature, see e.g., \cite{SAKRISON:1969165, Sakrison:1970, Poor:1982, Vikrant:2024}. 
However, although novel extensions of the classical rate-distortion have been introduced for various applications, the study of their distributionally robust counterparts has not received much attention. For example, neither the rate-distortion-perception function (RDPF) \cite{blau:2019}, describing the trade-off between compression rate, distortion fidelity, and perceptual quality of the compressed samples, nor the information rate function (IRF) \cite{theis:2021}, a general description of information-theoretic functionals considering per-letter constraints, has been extended to the robust setting. As shown in \cite{theis:2021}, both problems have operational significance when considering an encoding-decoding scheme with access to a (possibly infinite) source of common randomness. The proposed coding theorems leverage the common randomness to construct coding schemes based on the so-called \textit{SFRL} \cite{Li:2018}. However, the extension to the robust setting of this code construction has not yet been investigated.

In this paper, we propose a distributional robust formulation of the IRF, discussing both coding and computational aspects of the problem. To this end, we first propose two novel extensions of the SFRL. For a fixed conditional distribution $Q_{Y|X}$, we consider the couplings $(X,Y)$ induced by $X \sim \mu_X$ with $\mu_X$ belonging to a set $\mathcal{S}$ of possible distributions. Considering the cases where $\mathcal{S}$ is either a finite set of distributions (Theorem \ref{theorem:RSFRL1}) or a Kullback–Leibler divergence sphere centered at a fixed nominal distribution (Theorem \ref{theorem:RSFRL2}), we show that, for any $\mu_X \in \mathcal{S}$, $Y$ can be expressed as a deterministic function of $X$ and an auxiliary random variable $Z$. Then, considering the operational setup in \cite{theis:2021}, we use the derived robust SFRL extensions to characterize distributionally robust coding schemes for both the one-shot (Theorem \ref{theorem:OneShotIRF}) and asymptotic regime (Theorem \ref{theo:CodingTheoremDiscr}), generalizing previous results present in the literature \cite{SAKRISON:1969165,theis:2021}. Focusing on the R-RDF defined on abstract spaces, we then consider the specific case where $\mu_X$ belongs to a given Kullback–Leibler (KL)-sphere. We first show that, in this setting, the R-RDF is equivalent to a more tractable max-min problem (Theorem \ref{theorem:VonNeumann}), which allows us to derive an implicit characterization of the points attaining the R-RDF curve (Theorem \ref{theo:SolutionsMaxMin}). The derived solutions are later used to implement a novel algorithm for the computation of the R-RDF in the specific case of finite alphabet sources (Algorithm \ref{alg:RDFEstimation}). We corroborate our theoretical results using numerical simulations, comparing the estimation capabilities of the proposed algorithm against the closed form R-RDF derived for binary sources (Theorem \ref{theo:RRDFBern}).

\subsection{Notation}
We denote with $\mathbb{R}$ the set of real numbers and with $\mathbb{R}^+_0$ the set $[0, \infty)$. $\mathbb{N}$ denotes the set of natural numbers, while, for $a, b \in \mathbb{N}$, $a:b$ indicates the interval $[a, b] \subset \mathbb{N}$. Given a measurable space $(\mathcal{X}, \Sigma_{\mathcal{X}})$, let $\mathcal{M}_1(\mathcal{X})$ be the set of probability measures on $\mathcal{X}$. For $\mu \in \mathcal{M}_1(\mathcal{X})$, a function $f \in L_p(\mu)$ if $f$ is a real-valued measurable function and $(\int_\mathcal{X} |f|^pd\mu )^\frac{1}{p} < \infty$. Given a second measurable space $(\mathcal{Y},\Sigma_{\mathcal{Y}})$, we indicate the set of conditional distributions with $\mathcal{L}(\mathcal{X} \times \mathcal{Y}) \triangleq \{ Q~|~Q: \mathcal{X} \to \mathcal{M}_1(\mathcal{Y})~ \text{s.t.} ~ \forall A \in \Sigma_{\mathcal{Y}}, ~Q(A|\cdot) ~\text{is} ~ \text{measurable}\}$. Given $\mu_X \in \mathcal{M}_1(\mathcal{X})$ and a conditional measure $Q_{Y|X} \in \mathcal{L}(\mathcal{X} \times \mathcal{Y})$ describing the coupling $(X,Y) \sim \mu_X \otimes Q_{Y|X}$, we indicate their mutual information as either $I(X,Y)$ or $I(\mu_X, Q_{Y|X})$, depending on the context. Lastly, given $\mathcal{S} \subset \mathcal{M}_1(\mathcal{X})$ and a conditional measure $Q$, we denote $\bar{I}(\mathcal{S},Q) = \sup_{\mu \in \mathcal{S}} I(\mu,Q)$. 

\section{Preliminaries}
In this section, we provide the main assumptions regarding our setup, as well as the operational definitions that identify "good" distributionally robust source codes.
Although not strictly necessary, we assume throughout this paper that the source alphabet $\mathcal{X}$ and the reconstruction alphabet $\mathcal{Y}$ are Polish spaces. This choice stems from the observation that most source-coding applications consider sources defined over metrizable spaces.
In the following, we begin by defining the source code.
    
\begin{definition}\textit{(Source code)}
For sets $\mathcal{X}$ and $\mathcal{Y}$, we define a (possibly stochastic) $N$-block encoder as any function in $\mathcal{F}_N = \{f: \mathcal{X}^N \times \mathcal{Z} \to \mathbb{N}_0\}$, where $\mathcal{Z}$ denotes the domain of the auxiliary source of randomness. Similarly, a (stochastic) $N$-block decoder is a function in $\mathcal{G}_N = \{g:\mathbb{N}_0 \times \mathcal{Z} \to \mathcal{Y}^N\}$. An element of $\mathcal{F}_N \times \mathcal{G}_N$ is called an $N$-block code.
\end{definition}

In this paper, we focus on the case of \textit{distributionally robust source coding}, i.e., the statistics of the information source are not perfectly known a priori, but are assumed to belong to a set of possible sources $\mathcal{S}$. Therefore, in what follows, we define \textit{achievable} codes, differentiating between the one-shot and block coding settings. 
\begin{definition}\textit{(One-Shot Achievability)}
    Given a set of sources $\mathcal{S}$ and a set of fidelity constraints $\{h_j(P_{X,Y})\le \theta_j\}$, the rate $R$ is \textit{robust one-shot achievable} if there exists an encoder $f \in \mathcal{F}_1$, a decoder $g \in \mathcal{G}_1$ and a random variable $Z$ valued on $\mathcal{Z}$ such that, for any $\mu_X \in \mathcal{S}$  defining the source $X \sim \mu_X$,
    \begin{align*}
        M = f(X, Z) \quad \text{and} \quad Y = g(M, Z)
    \end{align*}
    guarantee that the conditional entropy $H(M|Z) \le R$ while the induced joint distribution $P_{X,Y}$ satisfies the given constraints.
\end{definition}

\begin{definition}\textit{(Asymptotic Achievability)}
    Given a set of sources $\mathcal{S}$ and a set of fidelity constraints $\{h_j(P_{X,Y})\le \theta_j\}$, the rate $R$ is \textit{asymptotically achievable} if there exists a sequence of codes $(f_N, g_N) \in \mathcal{F}_N \times \mathcal{G}_N$ and a random variable $Z$ valued in $\mathcal{Z}$ such that, for any $\mu_X \in \mathcal{S}$ inducing the independent and identically distributed sequence $X^N = (X_1, \ldots, X_N)$ with $ X_i \sim \mu_X$,
    \begin{align*}
        M = f_N(X^N, Z) \quad \text{and} \quad Y^N = g_N(M, Z)
    \end{align*}
    guarantee that, for $N$ large enough, the conditional entropy $\frac{H(M|Z)}{N} \le R$ while the induced joint distribution $P_{X_i,Y_i}$ satisfies the given constraint set for all $i \in \{ 1:N\}$.
\end{definition}

\section{Robust Strong Functional Representation Lemmas}
Our first technical results focus on the extension of the SFRL \cite{Li:2018} to the distributionally robust setting, where, for a fixed conditional distribution $Q$, the uncertainty about the true distribution of the coupling $(X,Y)$ stems from the uncertainty about the distribution of the marginal $X$, i.e., $\mu_X$ can be any of the distributions belonging to a set $\mathcal{S}$. We call this extension \textit{Robust SFRL} (R-SFRL), which will be the main tool for deriving the new robust source coding theorems in Section \ref{sec:rsct}.
\par With the following theorem, we begin by characterizing the case where $\mathcal{S}$ is a finite collection of distributions.

\begin{theorem}(Robust SFRL) \label{theorem:RSFRL1}
    Let $\mathcal{S} = \{ \mu_1,\ldots,\mu_{|\mathcal{S}|} \}$ be a finite set of distributions and, given a conditional distribution $Q$, let $(X_\alpha, Y_\alpha) \sim \mu_\alpha \otimes Q$ with $I(X_\alpha, Y_\alpha) < \infty$, for all $\alpha \in 1:|\mathcal{S}|$. Then, there exists a random variable $Z$ such that, for all $\alpha \in {1, \ldots, |\mathcal{S}|}$, $Z$ is independent of $X_\alpha$, i.e., $X_\alpha \perp Z$, and $Y_\alpha$ can be expressed as a deterministic function $g(X_{\alpha}, Z)$ of $(X_{\alpha}, Z)$ such that
    \begin{align}
        H(Y_{\alpha}|Z) \le \Imax(\mathcal{S},Q) + \log\left(|\mathcal{S}|\cdot(\Imax(\mathcal{S},Q) + 1)\right) + 4. \label{SFRL:main_eq}
    \end{align}
\end{theorem}
\begin{IEEEproof}
    See Appendix \ref{proof:RSFRL1}.
\end{IEEEproof}
A second interesting scenario is the case where $\mathcal{S}$ is a KL divergence sphere of radius $\Gamma$ centered in some nominal distribution $\mu_0$, where the belief on the nominal distribution being the real distribution is inversely proportional to $\Gamma$. The following theorem extends the SFRL to this case.

\begin{theorem} \label{theorem:RSFRL2}(KL-Robust SFRL)
Given a nominal distribution $\mu_0$ and $\Gamma \ge 0$, let $\mathcal{S}(\Gamma) = \{ \mu \in \mathcal{M}_1(\mathcal{X}): D_{KL}(\mu||\mu_0) \le \Gamma\}$. Assume that, for a given conditional distribution $Q$ and for all $\mu\in \mathcal{S}$ such that $(X, Y) \sim \mu \otimes Q$, $I(X, Y) < \infty$. Then, there exists a random variable $Z$ such that, for any $X \sim \mu$ with $\mu \in \mathcal{S}(\Gamma)$, $Z$ is independent of $X$ ($X \perp Z$) and $Y$ can be expressed as a deterministic function of $g(X, Z)$ of $(X,Z)$ such that
    \begin{align}
        H(Y|Z) \le \Imax(\mathcal{S},Q) + \Gamma + \log(\Imax(\mathcal{S},Q) +\Gamma +1) + 4.
    \end{align}
\end{theorem}
\begin{IEEEproof}
    See Appendix \ref{proof:RSFRL2}.
\end{IEEEproof}
We conclude this section with the following observation.
\begin{remark} (Special cases)
If, in Theorem \ref{theorem:RSFRL1} we consider a singleton set $\mathcal{S} = \{\mu_0\}$ or in Theorem \ref{theorem:RSFRL2} a KL-sphere with radius $\Gamma = 0$, then both Theorems \ref{theorem:RSFRL1} and \ref{theorem:RSFRL2} recover the original SFRL bound \cite{Li:2018}.
\end{remark}

\section{Robust Lossy Source Coding Theorems} \label{sec:rsct}
In this section, we study the application of R-SFRL to the robust source coding problem. To this end, we first introduce a robust version of the information rate function (IRF), originally described in \cite{theis:2021}.

\begin{definition}\textit{(Robust Information Rate Function (R-IRF))} \label{def:RIRF}
    Let $\mathcal{S} \subset \mathcal{M}_1(\mathcal{X})$ be a set of probability distributions, $\mathcal{H} = \{h_i\}_{i = 1}^M$ be a finite set of real-valued functions defined in the space $\mathcal{M}_1(\mathcal{X}\times\mathcal{Y})$ of joint distributions $P_{X,Y}$, and $\theta \in \mathbb{R}^M$. Then, the robust IRF is defined as
    \begin{align}
        \begin{split}
            \bar{R}(\theta) &\triangleq \inf_{Q_{Y|X} \in \mathcal{L}(\theta)} \Imax(S, Q_{Y|X}) \\
            &= \inf_{Q_{Y|X} \in \mathcal{L}(\theta)} \sup_{\mu_X \in \mathcal{S}} I(\mu_X, Q_{Y|X}) \label{eq:IRF}
        \end{split}
    \end{align}
    where the set $\mathcal{L}(\theta) \subseteq \mathcal{L}(\mathcal{X} \times\mathcal{Y})$ is defined as
    \begin{align}
        \mathcal{L}(\theta) \triangleq \left\{ Q_{Y|X}: ~h_i[\mu_X Q_{Y|X}] \le \theta_i ~, \forall \mu_X \in \mathcal{S}, i \in 1:M \right\}. \label{eq:stocahsticKer}
    \end{align}
\end{definition}

The definition of R-IRF is based on a min-max formulation of the IRF. For target constraint levels $\theta$, the uncertainty set $\mu_X \in \mathcal{S}$ can be seen as nature trying to maximize the rate, while the designer selects a code $Q_{Y|X} \in \mathcal{L}(\theta)$ aiming to minimize it.
The following remark elaborates on the generality of the R-IRF definition.

\begin{remark}
    Similarly to the IRF, the R-IRF provides a general description for a wide variety of information-theoretic quantities. For example, let $h_1$ and $h_2$ be defined as
    \begin{align}
        h_1(P_{X,Y}) = \mathbb{E}[\delta(X,Y)] \quad h_2(P_{X,Y}) = d(\mu_X, q_Y)
    \end{align}
    for appropriate choices of distortion metric $\delta$ and divergence function $d$. Then, the classical R-RDF \cite{SAKRISON:1969165} can be obtained by considering the set $\mathcal{H} = \{h_1\}$, while the set $\mathcal{H} = \{h_1, h_2\}$ can be seen as a novel robust extension of the RDPF.
\end{remark}

Due to its generality, the R-IRF is not guaranteed to admit a non-empty feasible solution for all constraint levels $\theta \in \mathbb{R}^M_+$. To facilitate our subsequent analysis, we will operate under the following assumption, leaving the investigation of the feasibility conditions of the R-IRF for future work.

\begin{assumption} \label{ass:feasibilityIRF}
We consider the set of possible constraint levels $\Theta = \{ \theta \in \mathbb{R}^M ~|~ \exists Q_{Y|X} \in \mathcal{L}(\theta) ~\text{s.t.} ~\Imax(S,Q_{Y|X}) < \infty\}$, i.e., we consider only $\theta$ for which $\mathcal{L}(\theta)$ has at least one feasible point. Hence, for $\theta \in \Theta$, the optimization problem \eqref{eq:IRF} defining the R-IRF is feasible, i.e., $\bar{R}(\theta) < \infty$. 
\end{assumption}

Under Assumption \ref{ass:feasibilityIRF}, we are now ready to present our achievability and converse results relative to the one-shot case.

\begin{theorem} \label{theorem:OneShotIRF}
Let $\mathcal{S} = \{ p_1,\ldots,p_{|\mathcal{S}|} \}$ be a finite set of distributions, $\mathcal{H} = \{h_i\}^M_i$ be a finite set of constraints and let $\theta$ the associated constraint levels. Then, $R$ is \textit{robust one-shot achievable} if 
\begin{align}
    R \ge \bar{R}(\theta) + \log\left(|S|\cdot(\bar{R}(\theta) + 1)\right) + 4
\end{align}
\end{theorem}
\begin{IEEEproof}
    See Appendix \ref{proof:OneShotIRF}.
\end{IEEEproof}

\begin{theorem} \label{theorem:OneShotIRFConverse}
Let $\mathcal{S} = \{ p_1,\ldots,p_{|\mathcal{S}|} \}$ be a finite set of distributions, $\mathcal{H} = \{h_i\}^M_i$ be a finite set of constraints and let $\theta$ the associated constraints levels. Then, $R$ is \textit{robust one-shot achievable} only if $R \ge \bar{R}(\theta)$.
\end{theorem}
\begin{IEEEproof}
    See Appendix \ref{proof:OneShotIRFConverse}.
\end{IEEEproof}

Unfortunately, the results of Theorems \ref{theorem:OneShotIRF} and \ref{theorem:OneShotIRFConverse} do not provide a tight characterization of the one-shot rate region. However, when extending the analysis to the asymptotic setting, the rate region becomes well characterized, as shown in the following theorem.

\begin{theorem} \label{theo:CodingTheoremDiscr} Let $\mathcal{S} = \{ p_1,\ldots,p_{|\mathcal{S}|} \}$ be a finite set of distributions, $\mathcal{H} = \{h_i\}^M_i$ be a set of constraints and let $\theta$ the associated constraints levels. Then, $R$ is asymptotically achievable if and only if $R \ge \bar{R}(\theta)$.
\end{theorem}
\begin{IEEEproof}
   See Appendix \ref{proof:CodingTheoremDiscr}.
\end{IEEEproof}
 
By leveraging variable-length coding with common randomness, Theorem \ref{theo:CodingTheoremDiscr} generalizes classical results in robust source coding to accommodate arbitrary sets of fidelity metrics $\mathcal{H}$. For example, Sakrison \cite{SAKRISON:1969165} characterized the same rate region using fixed-rate codes under an average single-letter distortion constraint, considering either finite or compact uncertainty sets $\mathcal{S}$. Given the similarities with \cite{SAKRISON:1969165}, we believe that a natural extension of Theorem \ref{theo:CodingTheoremDiscr} could consider source sets $\mathcal{S}$ that satisfy the same notion of compactness. However, this is left for future work.

\section{Case Study: Robust RDF}
In this section, we focus on a specific instance of the R-IRF, considering the classical problem of the distributionally robust RDF (R-RDF).
More specifically, we consider the case where the unknown true source probability measure is assumed to belong to the set
\begin{align*}
    \mathcal{S}(\Gamma) \triangleq \{ \mu \in \mathcal{M}_1(A): D_{KL}(\mu||\mu_0) \le \Gamma \}, \quad \Gamma \in [0, +\infty)
\end{align*}
i.e., a KL-sphere of radius $\Gamma$ centered at $\mu_0 \in \mathcal{M}_1(A)$. 

Given the set $\mathcal{S}(\Gamma) \subset \mathcal{M}_1(\mathcal{X})$ and a distortion function $\delta: \mathcal{X} \times \mathcal{Y} \to \mathbb{R}_0^+$, the set of stochastic kernels $\mathcal{L}(\theta)$ defined in \eqref{eq:stocahsticKer} becomes 
\begin{align}
    \mathcal{L}_{\mu}(D) &\triangleq \Big\{Q \in \mathcal{L}(\mathcal{X} \times \mathcal{Y}): 
    \int \delta(x,y) Q(dy|x)\mu(dx)  \le D \Big\} \nonumber\\
    \mathcal{L}(D) &\triangleq \bigcap_{\mu \in \mathcal{S}(\Gamma)} \mathcal{L}_{\mu}(D) \label{eq:ConstraintSetKern}
\end{align}
where $\mathcal{L}_{\mu}(D)$ is equivalent to the constraint set of the classical RDF for a fixed source $\mu$. Consequently, the constraint set of the R-RDF can be expressed as the intersection of the constraint set associated with each source in the considered class $\mathcal{S}(\Gamma)$. Therefore, following Definition \ref{def:RIRF} and using the described constraint set, the R-RDF is defined as
\begin{align}
    R^+(D) \triangleq \inf_{Q \in \mathcal{L}(D)} \sup_{\mu \in \mathcal{S}(\Gamma)} I(\mu, Q). \label{eq:rrdf+}
\end{align}
The min-max structure of \eqref{eq:rrdf+}, and of the R-IRF in general, can present numerous challenges in terms of computational complexity. For this reason, it is often preferable to consider an alternative max-min formulation of the problem, i.e., 
\begin{align}
    R^-(D) \triangleq  \sup_{\mu \in \mathcal{S}(\Gamma)} \inf_{Q \in \mathcal{L}(D)} I(\mu, Q) \label{eq:rrdf-}
\end{align}
which has the added advantage of sharing a similar structure to the classical RDF, i.e., the inner imfimum over $Q \in \mathcal{L}(D)$. However, in general, $R^+(D) \ge R^-(D)$, where equality holds if the conditions of the Min-Sup Theorem\footnote{Sometimes referred to also as Von Neumann condition \cite{sion:1958}.} hold for the considered problem \cite{sion:1958}. In the following theorem, we provide general conditions under which \eqref{eq:rrdf+} satisfies the min-sup conditions, i.e., \eqref{eq:rrdf+} is equivalent to \eqref{eq:rrdf-}. 

\begin{theorem} \label{theorem:VonNeumann}
Let $\delta: \mathcal{X} \times \mathcal{Y} \to \mathbb{R}_0^+$ be a measurable distortion function such that, for $\mu$-almost every $x \in \mathcal{X}$, the function $\delta(x,\cdot)$ is continuous on $\mathcal{Y}$. Then, there exists $Q^{*} \in \mathcal{L}(D)$ such that
    \begin{align*}
        R^-(D) = \sup_{\mu \in \mathcal{S}(\Gamma)} I(\mu, Q^*) = R^+(D)
    \end{align*}
\end{theorem}
\begin{IEEEproof}
See Appendix \ref{proof:VonNeumann}.
\end{IEEEproof}

Using the result of Theorem \ref{theorem:VonNeumann}, we devote the remainder of this section to characterizing the minimizers that achieve $R^{-}(D)$. To this end, we first recall a known result to characterize the conditional distribution that achieves the RDF for a fixed source on abstract spaces \cite{rezaei:2006rate}.

\begin{lemma} \label{lemma:RDsolution}
For a fixed $\mu \in \mathcal{S}(\Gamma)$ and any $s\geq{0}$, the inner infimum in \eqref{eq:rrdf-} is attained by $Q^* \in \mathcal{L}(D)$ given by
\begin{align}
Q^*(A|x) = \frac{\int_A e^{-s\delta(x,y)}q^*(dy)}{\int_{\mathcal{Y}} e^{-s\delta(x,y)}q^*(dy)} \label{eq:BestKernel}
\end{align}
where $A \subseteq \mathcal{Y}$ and $q^* \in \mathcal{M}_1(\mathcal{Y})$ is the marginal on $\mathcal{Y}$ of the measure $\mu \otimes Q$. Furthermore, the value of $\inf_{Q \in \mathcal{L}(D)} I(\mu, Q)$ is given by $F(q^*)$, where
    \begin{align*}
         F(q^*)= &\sup_{s \ge 0} \Bigg[-sD
        - \int_{\mathcal{A}} \log \left( \int_{\mathcal{Y}} e^{-s\delta(x,y)} q^*(dy) \right) \mu(dx) \Bigg].
    \end{align*}
\end{lemma}

\noindent From Lemma \ref{lemma:RDsolution}, \eqref{eq:rrdf-} can be expressed as
\begin{align}
    \begin{split}
        R^-(D) = & \sup_{s \ge 0} \sup_{\mu \in \mathcal{S}(\Gamma)} \Bigg[-sD \\
        &- \int_{\mathcal{A}} \log \left( \int_{\mathcal{Y}} e^{-s\delta(x,y)} q^*(dy) \right) \mu(dx) \Bigg]
    \end{split} \label{eq:FirstDual}
\end{align}
where the suprema over $\mu \in \mathcal{S}(\Gamma)$ and $s \ge 0$ are interchangeable. We can now introduce the Lagrangian functional for the constraint set $\mathcal{S}(\Gamma)$, i.e.,
\begin{align*}
    F_{s,\lambda}(\mu,q^*) \triangleq  
    \Big[&-sD -\lambda\left( D_{KL}(\mu||\mu_0) - \Gamma \right)\\
        &- \int_{\mathcal{A}} \log \left( \int_{\mathcal{Y}} e^{-s\delta(x,y)} q^*(dy) \right) \mu(dx) \Bigg]
\end{align*} 
where $\lambda \ge 0$ is the associated Lagrangian multiplier. 
Let $\mu^* = \arg \sup_{\mu \in \mathcal{S}(\Gamma)} F_{s,\lambda}(\mu,q^*)$. It can be shown, via Lagrange duality, that
\begin{align}
    R^-(D) = \eqref{eq:FirstDual} = \sup_{s\ge0} \inf_{\lambda \ge 0} F_{s,\lambda}(\mu^*,q^*). \label{eq:FinalDual}
\end{align}

\noindent In the next theorem, we use \eqref{eq:FinalDual} to provide a complete implicit characterization of the $R^-(D)$ problem.

\begin{theorem} \label{theo:SolutionsMaxMin}
    Assume $e^{\frac{g}{\lambda}} \in L_1(\mu_0)$ and $ge^{\frac{g}{\lambda}} \in L_1(\mu_0)$, where $g: \mathcal{X} \to \mathbb{R}$, $g(x) = -\log\left(\int_{\mathcal{Y}} e^{-s\delta(x,y)}q^*(dy) \right)$. Then, the solution to \eqref{eq:FinalDual} is given by
    \begin{align}
        R^-(D) &= sD + \lambda\Gamma \label{eq:optRmin}\\
        &+ \lambda \log\left( \int_\mathcal{X} \left( \int_\mathcal{Y} e^{-s\delta(x,y)} q^*(dy)  \right)^{-\frac{1}{\lambda}} \mu_0(dx) \right) \nonumber
    \end{align}
    where $s \ge 0, \lambda \ge 0 $ are, respectively, the Lagrangian multipliers associated with the distortion and uncertainty set constraints.
    Furthermore, the infimum over $Q \in \mathcal{L}(D)$ is attained by $\eqref{eq:BestKernel}$, while the supremum over the class of sources $\mu \in \mathcal{S}(\Gamma)$ is attained by
    \begin{align}
        \mu^*(dx) = \frac{\left( \int_{\mathcal{Y}} e^{-s\delta(x,y)}q^*(dy) \right)^{-\frac{1}{\lambda}} \mu_0(dx)}{ \int_\mathcal{X} \left(\int_{\mathcal{Y}} e^{-s\delta(x,y)}q^*(dy) \right)^{-\frac{1}{\lambda}} \mu_0(dx)}. \label{eq:bestSource}
    \end{align}
\end{theorem}
\begin{IEEEproof}
    See Appendix \ref{proof:SolutionsMaxMin}.
\end{IEEEproof}

We stress the following remark on the relation between the Lagrangian multipliers $(s, \lambda)$ and the constraint levels $(D, \Gamma)$. 
\begin{remark}
    Similarly to the classical RDF \cite{Csiszar:1974, blahut:1972computation}, assuming that that the Lagrangian multipliers $(s,\lambda)$ are fixed rathan than the constraint levels $(D,\Gamma)$ greatly simplifies the derivation of the optimal solutions $(\mu^*,Q^*)$. Although the existence of an inverse relationship can be proven between Lagrangian multipliers and constraint levels, the absence of an analytical mapping implies the necessity of search routines to identify the $(s,\lambda)$ inducing the specific levels $(D, \Gamma)$.
\end{remark}

\subsection{Computation of the Robust RDF for Discrete Sources}

We devote this section to applying the results of Theorem $\ref{theo:SolutionsMaxMin}$ for the computation of the R-RDF, focusing on the case of discrete sources, i.e., $\mathcal{X}$ and $\mathcal{Y}$ are finite sets. To this end, we exploit the fact that a pair $(\mu^*, Q^*)$ solving the system of equations \eqref{eq:BestKernel}-\eqref{eq:bestSource} is guaranteed to achieve $R^-(D)$. However, to simplify the derivation of an algorithmic solution, we first introduce an alternative to \eqref{eq:BestKernel} to characterize the optimal marginal distribution $q^*$. Considering the source $\mu^*$ fixed, it can be shown that the marginal $q^*$ must satisfy
\begin{align}
    c(y) = \sum_{x\in\mathcal{X}} \frac{e^{-s\delta(x,y)}}{\sum_{y \in \mathcal{Y}} e^{-s\delta(x,y)}q^*(dy)} \mu^*(dx) \le 1 ,\label{eq:CforMarginal}
\end{align}
which holds with equality $\forall y \in \mathcal{Y}$ such that $q^*(y) > 0$ \cite{blahut:1972computation}. Substituting \eqref{eq:bestSource} in \eqref{eq:CforMarginal}, we can obtain an implicit equation depending only on the marginal distribution $q^*$, i.e.,

\begin{align*}
    c(y) &= \frac{1}{Z} \sum_{x\in\mathcal{X}} \frac{e^{-s\delta(x,y)}}{\left(\sum_{y \in \mathcal{Y}} e^{-s\delta(x,y)}q^*(y) \right)^{1 + \frac{1}{\lambda}}} \mu_0(x)\\
    Z &= \sum_{x \in \mathcal{X}} \left(\sum_{y\in \mathcal{Y}} e^{-s\delta(x,y)}q^*(dy) \right)^{-\frac{1}{\lambda}} \mu_0(x).
\end{align*}

Unfortunately, verifying the inequality condition in \eqref{eq:CforMarginal}, therefore ensuring the optimality of $q^*$, remains a challenge from a computational standpoint. To partially solve this issue, we introduce the following simplified implicit function 
\begin{align}
    F:\mathcal{M}_1(\mathcal{Y}) \to \mathbb{R}^{|\mathcal{Y}|}, ~ F(q) = \left[ q(y)(1 - c(y)) \right]_{y \in \mathcal{Y}}. \label{eq:finalimplicit}
\end{align}
It is easy to check that $F(q) = 0$ represents a necessary optimality condition, implying that the set of roots of $F$ includes the optimal marginal $q^*$. The advantage of this formulation lies in the possibility of applying a root-finding method to identify a reduced set of candidate values $q$, which can then be tested for optimality using \eqref{eq:CforMarginal}. Algorithm \ref{alg:RDFEstimation} implements the ideas described so far, using Newton's method \cite{burden:2015numerical} as a root-finding routine. 

\begin{algorithm}[H]
    \caption{Discrete Robust RDF Estimation} \label{alg:RDFEstimation}
    \begin{algorithmic}[1]
        \Require Nominal distribution $\mu_0$; Set of initial points $\{q_i\}_{i=1}^N$; Lagrangian multipliers $s\ge0, \lambda\ge0$;
        Jacobian $J_F(\cdot)$; Error tolerance $\epsilon \ge 0$;
        \State $Op$ = \{\}
        \For{i = 1, \ldots, $N$}
            \Do \Comment{Newton's Method}
                \State $\eta \gets (J_F(q_i))^{-1} F(q_i)$  
                \State $q_i \gets q_i - \eta$ 
            \doWhile{$\eta \ge \epsilon$} 
            \If{$q_i$ satisfies \eqref{eq:CforMarginal}}
                \State $Op \gets \{q_i\} \cup Op$
            \EndIf
        \EndFor
        \Ensure Optimal Marginals $Op$.
    \end{algorithmic}
\end{algorithm}

\begin{remark} \label{remark:InitRandom}
    The implicit function $F$ can be shown to be non-monotonic for certain values of $(s,\lambda)$. In these cases, the convergence of Algorithm \ref{alg:RDFEstimation} heavily relies on the choice of the initialization point used in the root-finding routine.
    Although discretization-based schemes (e.g., \cite{wolfe:1993simple}) can guarantee the convergence of the algorithm, we empirically found that considering a single initial point, i.e., 
    \begin{align*}
        q(y) &\propto \sum_{x\in\mathcal{X}} \frac{e^{-s\delta(x,y)} (\mu_0(x))^{\frac{\lambda}{\lambda + 1}}}{\left(\sum_{y \in \mathcal{Y}} e^{-s\delta(x,y)} \right)^{\frac{\lambda + 1}{\lambda}}}
    \end{align*}
    yields good results in most cases, as shown in Section \ref{sec:numres}.
\end{remark}

To test the efficacy of Algorithm \ref{alg:RDFEstimation}, we use the following theorem as a benchmark, which provides the explicit characterization of the R-RDF under Hamming distortion for a class of Bernoulli sources.

\begin{theorem}\label{theo:RRDFBern}(R-RDF for Bernoulli sources) Let $\mathcal{X} = \{0,1\}$ and $\mathcal{S}(\Gamma) = \{ \mu \in \mathcal{M}_1(\mathcal{X}): D_{KL}(\mu|| \mu_0) \le \Gamma \}$ for some $\mu_0 = \text{Bern}(\beta_0)$. Then, the R-RDF $R^-(D)$ under Hamming distortion $\delta(x,y) = \mathbbm{1}_{(x \neq y)}$ is given by 
    \begin{align*}
        R^-(D) = h_b(\beta^*) - h_b(D), ~~\beta^* = \left( 1 + \left(\frac{1 - \beta_0}{\beta_0}\right)^{\frac{\lambda}{1 + \lambda}}\right)^{-1}
    \end{align*}
    where $h_b(
    \cdot)$ is the binary entropy function and $\lambda \ge 0$ is chosen such that $D_{KL}(\mu||\mu_0) \le \Gamma$ with $\mu = \text{Bern}(\beta^*)$.
    \begin{IEEEproof}
        See Appendix \ref{proof:RRDFBern}.
    \end{IEEEproof}
\end{theorem}

\section{Numerical Results} \label{sec:numres}
In this section, we provide empirical evidence of the performance of Algorithm \ref{alg:RDFEstimation} through numerical simulations. 

\begin{example}
Suppose $\mathcal{X} = \{0, 1\}$ and let $\mu_0 = Bern(0.1)$ be the nominal distribution. In Fig. \ref{fig:Rate-Distortion-Bernoulli}, we display a point estimate of $R^-(D)$ under Hamming distortion $\delta(x,y) = \mathbbm{1}_{(x \neq y)}$ resulting from Algorithm \ref{alg:RDFEstimation}, comparing it to the analytical expression recovered in Theorem \ref{theo:RRDFBern}. The resulting surface exhibits the expected convex-concave behavior with respect to $(D,\Gamma)$. Furthermore, we note that the estimate obtained from Algorithm \ref{alg:RDFEstimation} closely matches the analytical solution.

\begin{figure}[h]
    \centering
    \includegraphics[width=0.61\linewidth]{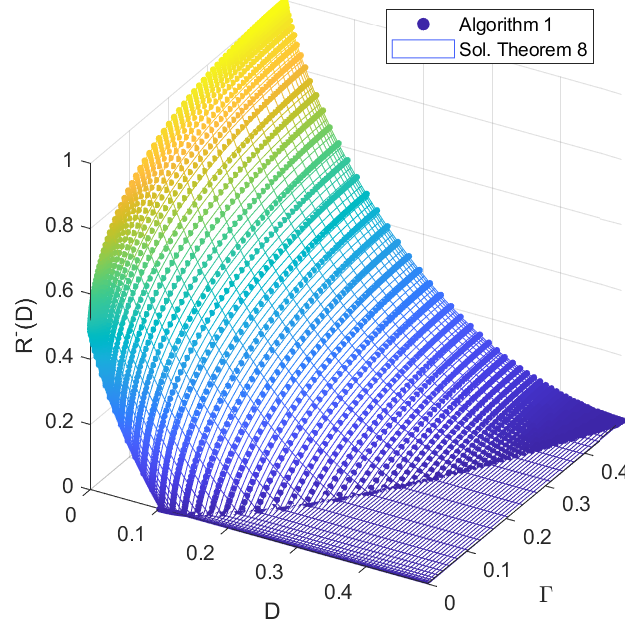}
    \caption{Robust RDF with nominal distribution $\mu_0 = Bern(0.1)$.} \label{fig:Rate-Distortion-Bernoulli}
\end{figure}
\end{example}

\begin{example}
Suppose $|\mathcal{X}| = 10$ and let
    \begin{align*}
        \mu_0 = &[0.0006, 0.0003, 0.09, 0.4262, 0.0002,\\& 0.1409, 0.0019, 0.0029, 0.3358, 0.0012]
    \end{align*}
be the given nominal distribution. Fig. \ref{fig:Rate-Distortion-N10} shows a point estimate of $R^-(D)$ under Hamming distortion resulting from Algorithm \ref{alg:RDFEstimation} along with the corresponding interpolated surface.

    \begin{figure}[h]
    \centering
    \includegraphics[width=0.61\linewidth]{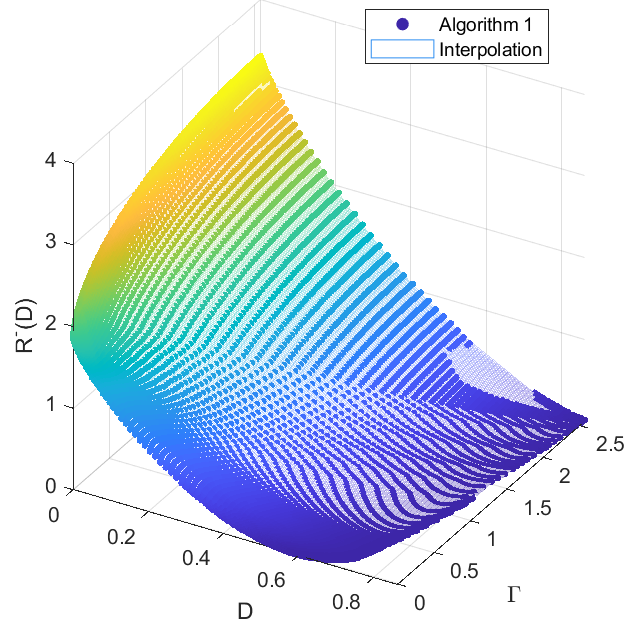}
    \caption{Robust RDF for with nominal distribution $\mu_0$.} \label{fig:Rate-Distortion-N10}
\end{figure}

We remark that, with the considered initialization conditions, Algorithm \ref{alg:RDFEstimation} fails to identify suitable solutions for larger values of $\Gamma$, i.e., $\lambda \to 0$. However, as pointed out in Remark \ref{remark:InitRandom}, testing a larger pool of initial points would solve this issue at the cost of increased computational complexity. 

\end{example}

\newpage

\IEEEtriggeratref{13}
\bibliographystyle{IEEEtran}
\bibliography{strings, biblio}

\clearpage

\appendices

\section{Proof of Theorem \ref{theorem:RSFRL1}}  \label{proof:RSFRL1}
To simplify the notation, for every $\mu_\alpha \in \mathcal{S}$ inducing $(X_\alpha, Y_\alpha) \sim \mu_\alpha \otimes Q$, we denote by $q_\alpha$ the associated marginal distribution of $Y_\alpha$. Furthermore, we denote by $\mathcal{S}^{'} = \{q_\alpha\}_{\alpha = 1}^{|S|}$ the set of such marginals. 
    \par Let $\Theta = \{\Theta_\alpha\}_{\alpha = 1}^{|\mathcal{S}|}$ where $\Theta_\alpha = \{(T_{\alpha,i}, \tilde{Y}_{\alpha,i})\}_{i = 1}^{\infty}$ is a non-homogeneous Poisson process (P.P.) with intensity measure $q_\alpha \times \lambda$, where $\lambda$ is the Lebesgue measure. For any fixed $x \in \mathcal{X}$ and any $\alpha = 1,\ldots,|S|$, define the mapping 
    \begin{align*}
        f_\alpha : (\tilde{y},t) \to \left(y, t \cdot \frac{dq_\alpha}{dQ(\cdot|x)}(\tilde{y})\right).
    \end{align*}
    Similarly to the construction of the SFRL \cite[Theorem 1]{Li:2018}, we apply the mapping $f_\alpha$ to the P.P. $\Theta_\alpha$, resulting in the point process
    \begin{align*}
        \left\{\left(\tilde{Y}_{\alpha,i}, T_{\alpha, i} \cdot \frac{dq_\alpha}{dQ(\cdot|x)}(\tilde{Y}_{\alpha,i})\right)\right\}
    \end{align*}
    which can be shown to be a P.P. with intensity measure $Q(\cdot|x) \times \lambda$. We define $\{\nu_\alpha\}$,$\{K_\alpha\}$, $K$, and $A$ as follows:
    \begin{align*}
        \nu_\alpha &= \min_i \left\{T_{\alpha, i} \cdot \frac{dq_\alpha}{dQ(\cdot|x)}(\tilde{Y}_{\alpha,i}) \right\}\\
        K_\alpha &= \arg\min_i \left\{T_{\alpha, i} \cdot \frac{dq_\alpha}{dQ(\cdot|x)}(\tilde{Y}_{\alpha,i}) \right\}\\
        K &= \min_\alpha K_\alpha\\
        A &= \argmin_\alpha K_\alpha
    \end{align*}
    We remark that, for all $\alpha$, $\nu_\alpha \sim \text{Exp}(1)$ and, conditioned on the realization of $\nu_\alpha$, $\tilde{Y}_{\alpha, K_\alpha} \sim Q(\cdot|x)$.
    \par Now, consider the following coding scheme. Assume that $\Theta$ is available to both the encoder and decoder as \textit{common randomness}, and let $M = (K,A)$ be the encoded variable. The decoder can isolate the P.P. associated with $A$ and output $Y = \tilde{Y}_{A, K} \sim Q(\cdot|x)$. Since $\tilde{Y}_{A, K}$ is a function of $(A,K)$ and the common randomness $\Theta$, we have that:
    \begin{align}
        H(Y|\Theta) \le H(K, A) \le H(K) + H(A). \label{theorem:SFRL1:eq1}
    \end{align}
    We are interested in bounding $H(Y|\Theta)$. Since $A$ takes values in the index set of possible source distributions $\mathcal{S}$, we can bound $H(A)$ as:
    \begin{align}
        H(A) \le \log(|\mathcal{S}|) \label{theorem:SFRL1:eq2}
    \end{align}
    On the other hand, we can derive the following
    \begin{align*}
        \mathbb{E}[\log(K)|X = x] &= \mathbb{E}\left[\min_{\alpha}\log(K_{\alpha}) | X = x \right] \\
        & \stackrel{(a)}{\le} \min_\alpha \mathbb{E}\left[\log(K_{\alpha}) | X = x \right] \\
        &\stackrel{(b)}{\le} \min_{q_\alpha \in \mathcal{S}^{'}}  D_{KL}( Q(\cdot|x) || q_{\alpha}) + c
    \end{align*}
    where $c = e^{-1}\log(e) + 1$, $(a)$ follows from Jensen's inequality, and $(b)$ leverages the results contained in the proof of \cite[Theorem 1]{Li:2018}. It follows that, for any $\mu_{\beta} \in \mathcal{S}$, it holds
    \begin{align}
        \mathbb{E}[\log(K)] & \le \mathbb{E}_{X\sim \mu_{\beta}}\left[\min_{q_\alpha \in \mathcal{S}^{'}}  D_{KL}( Q(\cdot|x) || q_{\alpha}) + c\right] \nonumber\\
        &\le \min_{q_\alpha \in \mathcal{S}^{'}} D_{KL}( \mu_{\beta} Q || \mu_{\beta} q_{\alpha} ) + c \nonumber\\
        &\stackrel{(c)}{\le} \max_{\mu_{\beta} \in \mathcal{S}} \min_{q_\alpha \in \mathcal{S}^{'}} D_{KL}( \mu_{\beta} Q(\cdot|X) || \mu_{\beta} q_{\alpha} ) + c \nonumber\\
        &\stackrel{(d)}{\le} \max_{\mu_{\beta}\in\mathcal{S} } I(\mu_{\beta},Q) + c \nonumber 
    \end{align}
    where $(c)$ follows from taking the maximum over all possible sources $\mu_{\beta} \in \mathcal{S}$ and $(d)$ follows from
    \begin{align}
        \begin{split}
            D_{KL}( \mu_{\beta} Q(\cdot|X) || \mu_{\beta} q_{\alpha} ) = I(\mu_{\beta},Q) + D_{KL}(q_{\beta}||q_\alpha)\\
            \Rightarrow{}
            \min_{q_\alpha} D_{KL}( \mu_{\beta} Q(\cdot|X) || \mu_{\beta} q_{\alpha} ) = I(\mu_{\beta},Q).
        \end{split} \label{theorem:SRFL1:eq4}
    \end{align}
Using the results in \cite[Appendix B]{Li:2018} which characterizes the maximum entropy distribution of $K$ under the constraint $\mathbb{E}[\log(K)] \le \Imax(\mathcal{S},Q) + c $, we obtain that
    \begin{align}
        H(K) \le \Imax(\mathcal{S},Q) + \log(\Imax(\mathcal{S},Q) +1) + 4 \label{theorem:SFRL1:eq6}
    \end{align}
    Substituting \eqref{theorem:SFRL1:eq2} and \eqref{theorem:SFRL1:eq6} in \eqref{theorem:SFRL1:eq1}, we recover \eqref{SFRL:main_eq}. This concludes the proof.

\section{Proof of Theorem \ref{theorem:RSFRL2}} \label{proof:RSFRL2}
Similarly to the proof of Theorem \ref{theorem:RSFRL1}, for every $\mu \in \mathcal{S}$ inducing $(X, Y) \sim \mu \otimes Q$, we will indicate with $q$ the marginal distribution associated with $Y$, while $\mathcal{S}^{'}$ identifies the set of such marginals. Furthermore, let $q_0$ indicate the marginal distribution associated with the nominal distribution $\mu_0$. 
\par Let $\Theta = \{(T_{i}, \tilde{Y}_{0,i})\}_{i = 1}^{\infty}$ be a non-homogeneous Poisson process (P.P.) with intensity measure $q_0 \times \lambda$, where $\lambda$ is the Lebesgue measure. For any fixed $x \in \mathcal{X}$, define the mapping 
\begin{align*}
    f : (\tilde{y},t) \to \left(y, t \cdot \frac{dq_0}{dQ(\cdot|x)}(\tilde{y})\right).
\end{align*}
Similarly to the proof of \cite[Theorem 1]{Li:2018}, we apply the mapping $f_\alpha$ to the P.P. $\Theta_\alpha$, resulting in the point process
\begin{align*}
    \left\{\left(\tilde{Y}_{0,i}, T_{0, i} \cdot \frac{dq_0}{dQ(\cdot|x)}(\tilde{Y}_{0,i})\right)\right\}
\end{align*}
which can be shown to be a P.P. with intensity measure $Q(\cdot|x) \times \lambda$. Let $\nu$ and $K$ be defined as:
\begin{align*}
    \nu &= \min_i \left\{T_{0, i} \cdot \frac{dq_0}{dQ(\cdot|x)}(\tilde{Y}_{0,i}) \right\}\\
    K &= \arg\min_i \left\{T_{0, i} \cdot \frac{dq_0}{dQ(\cdot|x)}(\tilde{Y}_{0,i}) \right\}\\
\end{align*}
We note that, by the properties of the mapped P.P., $\nu \sim \text{Exp}(1)$ and, conditioned on the realization of $\nu$, $\tilde{Y}_{0,K} \sim Q(\cdot|x)$.
Following the approach in \cite[Theorem 1]{Li:2018}, we consider a coding scheme in which the common randomness $\Theta$ is available at both the encoder and decoder. The encoder transmits the index $K$, and the decoder, using $\Theta$, reconstructs the output as $Y = \tilde{Y}{0,K}$. Since $Y$ is a function of $K$ and the common randomness $\Theta$, it follows that $H(Y|\Theta) \le H(K)$.\\
From the results of \cite[Theorem 1]{Li:2018}, we have that
\begin{align*}
    \mathbb{E}[\log(K)|X = x] \le  D_{KL}( Q(\cdot|x) || q_{0}) + c
\end{align*}
where $c = e^{-1}\log(e)$. Hence, for any $\mu \in \mathcal{S}$
\begin{align*}
    \mathbb{E}[\log(K)] &\le D_{KL}( \mu \otimes Q || \mu \times q_{0}) + c\\
    &\le \max_{\mu \in \mathcal{S}} D_{KL}( \mu \otimes Q || \mu \times q_{0}) + c\\
    &\stackrel{(a)}{=} \max_{\mu \in \mathcal{S}} I(\mu, Q) + D_{KL}(q||q_0) + c\\
    &\stackrel{(b)}{\le} \max_{\mu \in \mathcal{S}} I(\mu, Q) + \Gamma + c
\end{align*}
where $(a)$ results from \eqref{theorem:SRFL1:eq4} and $(b)$ stems from the data processing inequality, i.e., $D_{KL}(q||q_0) \le D_{KL}(\mu||\mu_0) \le \Gamma$.
Leveraging the results in \cite[Appendix B]{Li:2018} characterizing the maximum entropy distribution of $K$ under the constraint $\mathbb{E}[\log(K)] \le \Imax(\mathcal{S},Q) + \Gamma + c $, we obtain that
\begin{align*}
    H(K) \le \Imax(\mathcal{S},Q) + \Gamma + \log(\Imax(\mathcal{S},Q) +\Gamma +1) + 4, \label{theorem:SFRL1:eq5}.
\end{align*}
This concludes the proof.

\section{Proof of Theorem \ref{theorem:OneShotIRF}} \label{proof:OneShotIRF}

The proof follows from the results of Theorem \ref{theorem:RSFRL1}. Under Assumption \ref{ass:feasibilityIRF}, the IRF is feasible and thus there exists a $Q_{Y|X}$ such that 
\begin{align*}
    \forall i: h_i(P_{X,Y}) \le \theta_i, ~ \text{and} ~ \Imax(S,Q) \le \bar{R}(\theta) + \epsilon
\end{align*}
for $\epsilon > 0$. Following the construction of random variables $(A,K)$ detailed in the proof of Theorem \ref{theorem:RSFRL1}, we have that
\begin{align*}
    H(A, K) \le \bar{R}(\theta) + \log\left(|S|\cdot(\bar{R}(\theta) + \epsilon + 1)\right) + 4 + \epsilon \le R
\end{align*}
for $\epsilon \to 0$. Since in our encoding scheme $M = (A,K)$, it follows that
\begin{align*}
    H(M|\Theta) \le H(M) = H(A, K) \le R,
\end{align*}
thus concluding the proof.

\section{Proof of Theorem \ref{theorem:OneShotIRFConverse}} \label{proof:OneShotIRFConverse}

The converse theorem follows the same structure of \cite[Theorem 2]{theis:2021}, observing that the rate $R$ must satisfy the worst-case scenario among all possible source distributions $\mu_X \in \mathcal{S}$, i.e., 
\begin{align*}
    R &\ge \max_{\mu_X \in \mathcal{S}} H(f(X,Z)|Z) \\
    &\ge \max_{\mu_X \in \mathcal{S}} I(X,f(X,Z)|Z)\\
    &\ge \max_{\mu_X \in \mathcal{S}} I(X,f(X,Z)|Z) + I(X,Z)\\
    &\ge \max_{\mu_X \in \mathcal{S}} I(X,(Y,Z)) \ge \max_{\mu_X \in \mathcal{S}} I(X,Y) \ge \bar{R}(\theta)
\end{align*}
This concludes the proof.

\section{Proof of Theorem \ref{theo:CodingTheoremDiscr}} \label{proof:CodingTheoremDiscr}

 Under Assumption \ref{ass:feasibilityIRF}, there exists a $Q_{Y|X}$ such that $Q_{Y,X} \in \mathcal{L}(\theta)$ and  $\Imax(\mathcal{S}, Q_{Y,X}) < \bar{R}(\theta) + \frac{1}{N}$. To leverage the results of Theorem \ref{theorem:RSFRL1}, we can construct the conditional distribution as the product measure
    \begin{align*}
        Q_{Y^N|X^N} = \prod_{i = i}^N Q_{Y|X}
    \end{align*}
    guaranteeing that, for all $\mu_X \in \mathcal{S}$, $(X_n, Y_n) \sim \mu_X \otimes Q_{Y|X}$ satisfies the IRF constraints. Furthermore, the product structure implies
    \begin{align}
        I(X^N,Y^N) \le N \Imax(\mathcal{S}, Q_{Y|X}).
    \end{align}
    It follows that, similarly to Theorem \ref{theorem:OneShotIRF}, the average entropy of the message $M = (K_N, A_N)$ satisfies
    \begin{align*}
        \frac{H(M)}{N} &\le \frac{1}{N}\big( N\Imax(\mathcal{S}, Q_{Y|X}) \\
        & \qquad + \log\left(|S|\cdot(N\Imax(\mathcal{S}, Q_{Y|X}) + 1)\right) + 4\big)\\
        &\le \bar{R}(\theta) + \frac{1}{N}\log\left(|S|\cdot(N\bar{R}(\theta) + 2)\right) + \frac{5}{N}
    \end{align*}
    Therefore, for $N \to \infty$, $\frac{H(M)}{N} \to \bar{R}(\theta)$, proving the achievability of the coding scheme. 
    \par The converse of the coding theorem can be recovered using a similar strategy to that in the proof of Theorem \ref{theorem:OneShotIRFConverse}, i.e.,
    \begin{align*}
        R &\ge \lim_{N\to\infty}\max_{\mu_X \in \mathcal{S}} \frac{H(f_N(X^N|Z))}{N} \\
        &\ge \lim_{N\to\infty}\max_{\mu_X \in \mathcal{S}} \frac{I(X^N,Y^N)}{N}\\
        &\ge \max_{\mu_X \in \mathcal{S}} I(X,Y) \ge \bar{R}(\theta)
    \end{align*}
    since it can be observed that, for all $N$,
    \begin{align*}
        I(X^N,Y^N) &= h(X^N) - H(X^N|Y^N)\\
                   &\stackrel{(a)}{\ge} \sum_{i = 1}^N h(X_i) - h(X_i|Y^N, X_1, \ldots, X_{i-1})\\
                   &\stackrel{(b)}{\ge} \sum_{i = 1}^N  h(X_i) - h(X_i|Y_i)\\
                   &\ge \sum_{i = 1}^N  I(X_i,Y_i) = N I(X,Y)
    \end{align*}
    where $(a)$ derives from the fact that $X^N$ is a sequence of independent symbols and $(b)$ derives from the properties of conditional entropy. This concludes the proof.

\section{Proof of Theorem \ref{theorem:VonNeumann}}\label{proof:VonNeumann}

To provide a complete and general proof, we require the introduction of appropriate topologies and function spaces on which the functions $R^+$ and $R^-$ are evaluated. We recall that the spaces $\mathcal{X}$ and $\mathcal{Y}$ are assumed to be Polish spaces.
\par Let $BC(\mathcal{Y})$ denote the vector space of bounded and continuous real-valued functions defined on $\mathcal{Y}$. Equipped with the sup norm topology, $(BC(\mathcal{Y}), ||\cdot||_{\infty}$) is a Banach space. Let $(BC(\mathcal{Y}))^*$ denote the topological dual of $BC(\mathcal{Y})$. Then, following \cite[Theorem IV.6.2]{dunford:1988}, $(BC(\mathcal{Y}))^*$ is isometrically isomorphic to the Banach space of finitely additive regular sign measures on $\mathcal{Y}$, i.e., $(BC(\mathcal{Y}))^* \simeq \mathcal{M}_{rba}(\mathcal{Y})$. Let $\Pi_{rba}(\mathcal{Y}) \subset \mathcal{M}_{rba}$ denote the set of regular bounded finitely additive probability measures on $\mathcal{Y}$.
For $\mu \in \mathcal{M}_1(\mathcal{X})$, we indicate with $L_1(\mu, BC(\mathcal{Y}))$ the space of all $\mu$-integrable functions defined on $\mathcal{X}$ and valued in $BC(\mathcal{Y})$. Equipping $L_1(\mu, BC(\mathcal{Y}))$ with the norm topology induced by the norm
\begin{align*}
   \phi \in  L_1(\mu, BC(\mathcal{Y}))~~ ||\phi||_{\mu} = \int_{\mathcal{X}} ||\phi(x)(\cdot)||_{BC(\mathcal{Y})} \mu(dx) < \infty
\end{align*}
defines the Banach space $\left(L_1(\mu, BC(\mathcal{Y})), ||\cdot||_{\mu}\right)$. Unfortunately, without any additional assumption, 
$(BC(\mathcal{Y}))^* \simeq \mathcal{M}_{rba}(\mathcal{Y})$ does not satisfy the Radon Nikodym property (RNP) \cite[Theorems III.5 and IV.1]{Diestel:1977}, implying that the dual of $L_1(\mu, BC(\mathcal{Y}))$ is not $L_\infty(\mu, \mathcal{M}_{rba}(\mathcal{Y}))$\footnote{Under the additional assumption of $\mathcal{Y}$ being a compact space, $(BC(\mathcal{Y}))^* \simeq \mathcal{M}_{ca}(\mathcal{Y})$ is the space of countably additive sign measures, which satisfy the RNP.}.
However, it follows from the theory of "lifting" \cite[Theorems VII.7 and VII.9]{tulcea:2012topics} that the dual of $L_1(\mu, BC(\mathcal{Y}))$ can be identified with $L^w_\infty(\mu, \mathcal{M}_{rba}(\mathcal{Y}))$, the space of all functions defined on $\mathcal{X}$ and valued on $\mathcal{M}_{rba}(\mathcal{Y})$ which are weak$^*$ measurable, i.e., $Q \in L^w_\infty(\mu, \mathcal{M}_{rba}(\mathcal{Y}))$ if, for each $\phi \in BC(\mathcal{Y})$, $x \to \int_{\mathcal{Y}} \phi(y)Q(dy; x)$ is $\mu$-measurable and $\mu$-essentially bounded.
Now, we define the following set
\begin{align*}
    \mathcal{L}_{ad} \triangleq L^w_\infty(\mu, \Pi_{rba}(\mathcal{Y})) \subset L^w_\infty(\mu, \mathcal{M}_{rba}(\mathcal{Y}))
\end{align*}
which represents the unit sphere in the space $L^w_\infty(\mu, \mathcal{M}_{rba}(\mathcal{Y}))$. Following Alaoglu's Theorem \cite[Theorem 2.6.18]{Megginson:2012}, $\mathcal{L}_{ad}$ is ensured to be weak$^*$ compact.
For each $\phi \in L_1(\mu, BC(\mathcal{Y}))$, we can define the linear functional on $L^w_\infty(\mu, \mathcal{M}_{rba}(\mathcal{Y}))$
\begin{align*}
    l_\phi(Q) \triangleq \int_{\mathcal{X}} \left( \int_{\mathcal{Y}} \phi(x, y) Q(dy; x) \right) \mu(dx)
\end{align*}
which is certainly bounded, linear, and weak$^*$-continuous. 
\par Under this function space formulation, we can consider a distortion function $\delta: \mathcal{X} \times \mathcal{Y} \to \mathbb{R}^+_0$ as a measurable function from the class $L_1(\mu, BC(\mathcal{Y}))$ subject to the constraint
\begin{align}
    \mathcal{L}_\mu(D) = \left\{ Q \in \mathcal{L}_{ad}: l_{\delta}(Q) \le D \right\}
\end{align}
As a subset of $\mathcal{L}_{ad}$ the set $\mathcal{L}_\mu(D)$ is convex, bounded, and weak$^*$-closed, hence weak$^*$-compact (being a closed subset of a weak$^*$ compact set).
\par The assumption on $\delta \in L_1(\mu, BC(\mathcal{Y}))$, i.e., a bounded distortion metric, can be relaxed by assuming that, for each $x \in \mathcal{X}$, $\delta(x, \cdot)$ is continuous on $\mathcal{Y}$. It can be shown that, in this case, $\mathcal{L}_\mu(D)$ remains weak$^*$-closed, i.e., weak$^*$-compact \cite[Lemma 2.4]{rezaei:2006rate}.

Given the general characterization of the properties of $\mathcal{L}_{\mu}(D)$ for any $\mu \in \mathcal{S}(\Gamma)$, the following lemma investigates the properties of the intersection of such sets. 
\begin{lemma} \label{lemma:weakstarcompact}
    Any family of intersections of $\mathcal{L}_{\mu}(D)$, for $\mu \in \mathcal{S}(\Gamma)$, is convex and weak$^*$-compact. 
\end{lemma}
\begin{IEEEproof}
    Regarding the convexity of the intersection, see \cite[Proposition 1.1.(4)]{Megginson:2012}. Regarding compactness, by \cite[Proposition B.10] {Megginson:2012}, the arbitrary intersection of weak$^*$-closed sets is a weak$^*$-closed set. Since the intersection is a subset of $\mathcal{L}_{ad}$, which is itself a weak$^*$-compact set, it follows that the intersection is also weak$^*$-compact. 
\end{IEEEproof}

We are now ready to prove the theorem. Let $\mathcal{A} = \mathcal{L}(D)$  and $\mathcal{B} = \mathcal{S}(\Gamma)$. By Lemma \ref{lemma:weakstarcompact}, the set $\mathcal{A}$ is convex and weak$^*$-compact. The set $\mathcal{B}$ is convex as well, due to the convexity of the KL divergence. Moreover, $I(\mu, Q)$ is convex and lower semi-continuous in $Q \in \mathcal{A}$, for every $\mu \in \mathcal{B}$, and concave in $\mu \in \mathcal{B}$, for every $Q \in \mathcal{A}$ \cite{Csiszar:1974}. Therefore, the conditions of the Min-Sup Theorem \cite{sion:1958} are satisfied. 

\section{Proof of Theorem \ref{theo:SolutionsMaxMin}} \label{proof:SolutionsMaxMin}

For fixed $(s, \lambda)$, $F_{s,\lambda}(\mu^*, q^*)$ can be expressed as
\begin{align*}
    F_{s,\lambda}(\mu^*, q^*) &= \sup_{\mu \in \mathcal{M}_1(\mathcal{X})}F_{s,\lambda}(\mu, q^*)\\
    &= - sD + \lambda\Gamma \\
    & \quad + \lambda\left(\sup_{\mu \in \mathcal{M}_1(\mathcal{X})} \int_{\mathcal{X}} \frac{g(x)}{\lambda} d\mu -D_{KL}(\mu||\mu_0)  \right)
\end{align*}
Since $s \ge 0$, it follows that $g(\cdot) \ge 0$, implying that $g$ is a measurable function bounded from below on $(\mathcal{X}, \Sigma_{\mathcal{X}})$. Applying the duality between relative entropy and free energy \cite[Proposition 2.3]{Pra:1996}, the supremum is achieved by $\mu^*$ characterized in \eqref{eq:bestSource}, which, once substituted in \eqref{eq:FinalDual}, recovers the expression \eqref{eq:optRmin}. This concludes the proof.

\section{Proof of Theorem \ref{theo:RRDFBern}} \label{proof:RRDFBern}

Following \eqref{eq:ConstraintSetKern}, we have $\mathcal{L}(D) \subset \bigcup_{\mu \in {\mathcal{S}}(\Gamma)} \mathcal{L}_\mu(D)$. Therefore, from classical results on the RDF of Bernoulli sources \cite[Theorem 10.3.1]{cover:1999elements}, we have
        \begin{align}
            R^{-}(D) \ge \max_{ \beta \in \mathcal{B}} h_b(\beta) - h_b(D) \label{eq:Bern_eq1}
        \end{align}
        where $\mathcal{B} = \{ \beta \in [0,1]: Bern(\beta) \in \mathcal{S}(\Gamma)\}$. We remark that the maximization problem over the source parameter $\beta$ is convex. Therefore, introducing the Lagrangian multiplier $\lambda \ge 0$ and solving the Karush-Kuhn-Tucker conditions \cite{rockafellar:1997convex}, we can characterize the sufficient optimality condition
        \begin{align*}
            \frac{1 - \beta^{*}}{\beta^{*}} = \left( \frac{1-\beta_0}{\beta_0} \right)^{\frac{\lambda}{1 + \lambda}}.
        \end{align*}
        To prove that \eqref{eq:Bern_eq1} holds with equality, we just need to show that the conditional distribution $Q_{\beta^*}$ achieving the RDF for $X \sim Bern(\beta^*)$, i.e., 
        \begin{align*}
            Q_{\beta^*}(1|0) &= \frac{D}{1-\beta^*}\frac{\beta^* - D}{1 - 2D}\\ Q_{\beta^*}(0|1) &= \frac{D}{\beta^*}\frac{(1 -\beta^*) - D}{1 - 2D}
        \end{align*}
        belongs to the set $\mathcal{L}(D)$, i.e., for all $\mu \in \mathcal{S}(\Gamma)$, 
        \begin{align}
            \mathbb{E}_{\mu Q_{\beta^*}}[\delta(X,Y)] = \beta Q_{\beta^*}(0|1) + (1 - \beta) Q_{\beta^*}(1|0) \le D. \label{eq:Bern_eq2}
        \end{align}
        Considering the limits $\lambda \to 0$ and $\lambda \to \infty$, we remark that $\beta^* \in [0,0.5]$. Therefore, $Q_{\beta^*}(0|1) \ge Q_{\beta^*}(0|1)$. 
        \par Considering the case $\mathcal{B} \subset [0,0.5)$, the monotonicity of $h_b(\cdot)$ implies that $\beta^* = \max \mathcal{B}$. Since $\mathbb{E}_{\mu Q_{\beta^*}}[\delta(X,Y)]$ is monotonically increasing over $[0,0.5)$, \eqref{eq:Bern_eq2} is satisfied.
        In all other cases, $\beta^* = 0.5$, implying that $Q_{\beta^*}(0|1) = Q_{\beta^*}(0|1) = D$, thus automatically satisfying \eqref{eq:Bern_eq2}.
        This concludes the proof.

\end{document}